\documentclass[twocolumn]{aastex62}

\newcommand\apjcls{1}
\newcommand\aastexcls{2}
\newcommand\othercls{3}

\newcommand\papercls{\aastexcls}
\if\papercls \apjcls
\usepackage{apjfonts}
\else\if\papercls \othercls
\usepackage{epsfig}
\usepackage{margin}
\usepackage{times}
\fi\fi
\usepackage{ifthen}
\usepackage{natbib}
\usepackage{bm}
\usepackage{amssymb, amsmath}
\usepackage{appendix}
\usepackage{etoolbox}
\usepackage[T1]{fontenc}
\usepackage{paralist}
\usepackage{newtxtext,newtxmath}
\if\papercls \apjcls
\newcommand\aas{\ref@jnl{AAS Meeting Abstracts}}
\newcommand\dps{\ref@jnl{AAS/DPS Meeting Abstracts}}
\newcommand\maps{\ref@jnl{MAPS}}
\else\if\papercls \othercls
\usepackage{astjnlabbrev-jh}
\fi\fi

\if\papercls \aastexcls
\hypersetup{citecolor=blue, 
            linkcolor=blue, 
            menucolor=blue, 
            urlcolor=blue}  
\else
\usepackage[
bookmarks=true,           
bookmarksnumbered=true,   
colorlinks=true,          
citecolor=blue,           
linkcolor=blue,           
menucolor=blue,           
urlcolor=blue,            
linkbordercolor={0 0 1},  
pdfborder={0 0 1},
frenchlinks=true]{hyperref}
\fi
\if\papercls \othercls

\else

\fi

\providecommand{\adsurl}[1]{\href{#1}{ADS}}

\makeatletter
\patchcmd{\NAT@citex}
  {\@citea\NAT@hyper@{%
     \NAT@nmfmt{\NAT@nm}%
     \hyper@natlinkbreak{\NAT@aysep\NAT@spacechar}{\@citeb\@extra@b@citeb}%
     \NAT@date}}
  {\@citea\NAT@nmfmt{\NAT@nm}%
   \NAT@aysep\NAT@spacechar\NAT@hyper@{\NAT@date}}{}{}

\patchcmd{\NAT@citex}
  {\@citea\NAT@hyper@{%
     \NAT@nmfmt{\NAT@nm}%
     \hyper@natlinkbreak{\NAT@spacechar\NAT@@open\if*#1*\else#1\NAT@spacechar\fi}%
       {\@citeb\@extra@b@citeb}%
     \NAT@date}}
  {\@citea\NAT@nmfmt{\NAT@nm}%
   \NAT@spacechar\NAT@@open\if*#1*\else#1\NAT@spacechar\fi\NAT@hyper@{\NAT@date}}
  {}{}
\makeatother

\makeatletter
\DeclareRobustCommand{\lowcase}[1]{\@lowcase#1\@nil}
\def\@lowcase#1\@nil{\if\relax#1\relax\else\MakeLowercase{#1}\fi}
\makeatother

\DeclareSymbolFont{UPM}{U}{eur}{m}{n}
\DeclareMathSymbol{\umu}{0}{UPM}{"16}
\let\oldumu=\umu
\renewcommand\umu{\ifmmode\oldumu\else\math{\oldumu}\fi}

\if\papercls \othercls

\else

\fi

\let\oldsim=\sim
\renewcommand\sim{\ifmmode\oldsim\else\math{\oldsim}\fi}
\let\oldpm=\pm
\renewcommand\pm{\ifmmode\oldpm\else\math{\oldpm}\fi}
\newcommand\by{\ifmmode\times\else\math{\times}\fi}

\newbox{\wdbox}
\renewcommand\c{\setbox\wdbox=\hbox{,}\hspace{\wd\wdbox}}
\renewcommand\i{\setbox\wdbox=\hbox{i}\hspace{\wd\wdbox}}

\newcount\timect
\newcount\hourct
\newcount\minct
\newcommand\now{\timect=\time \divide\timect by 60
         \hourct=\timect Cltiply\hourct by 60
         \minct=\time \advance\minct by -\hourct
         \number\timect:\ifnum \minct < 10 0\fi\number\minct}

\catcode`@=11

\newcommand\comment[1]{}

\newcommand\commenton{\catcode`\%=14}

\renewcommand\math[1]{$#1$}
\newcommand\mathshifton{\catcode`\$=3}

\let\atab=&
\newcommand\atabon{\catcode`\&=4}

\let\oldmsp=\sp
\let\oldmsb=\sb
\def\sp#1{\ifmmode
           \oldmsp{#1}%
         \else\strut\raise.85ex\hbox{\scriptsize #1}\fi}
\def\sb#1{\ifmmode
           \oldmsb{#1}%
         \else\strut\raise-.54ex\hbox{\scriptsize #1}\fi}
\newbox\@sp
\newbox\@sb
\def\sbp#1#2{\ifmmode%
           \oldmsb{#1}\oldmsp{#2}%
         \else
           \setbox\@sb=\hbox{\sb{#1}}%
           \setbox\@sp=\hbox{\sp{#2}}%
           \rlap{\copy\@sb}\copy\@sp
           \ifdim \wd\@sb >\wd\@sp
             \hskip -\wd\@sp \hskip \wd\@sb
           \fi
        \fi}
\def\msp#1{\ifmmode
           \oldmsp{#1}
         \else \math{\oldmsp{#1}}\fi}
\def\msb#1{\ifmmode
           \oldmsb{#1}
         \else \math{\oldmsb{#1}}\fi}

\def\supon{\catcode`\^=7}

\def\subon{\catcode`\_=8}

\def\supsubon{\supon \subon}

\newcommand\actcharon{\catcode`\~=13}

\newcommand\paramon{\catcode`\#=6}

\comment{And now to turn us totally on and off...}

\newcommand\reservedcharson{ \commenton  \mathshifton  \atabon  \supsubon 
                             \actcharon  \paramon}

\catcode`@=12
\reservedcharson

\if\papercls \apjcls

\else

\fi

\newcommand\chisq{\ifmmode{\chi\sp{2}}\else\math{\chi\sp{2}}\fi}
\newcommand\redchisq{\ifmmode{ \chi\sp{2}\sb{\rm red}}
                    \else\math{\chi\sp{2}\sb{\rm red}}\fi}
\newcommand\Teq{\ifmmode{T\sb{\rm eq}}\else$T$\sb{eq}\fi}
\newcommand\mjup{\ifmmode{M\sb{\rm Jup}}\else$M$\sb{Jup}\fi}
\newcommand\rjup{\ifmmode{R\sb{\rm Jup}}\else$R$\sb{Jup}\fi}
\newcommand\msun{\ifmmode{M\sb{\odot}}\else$M\sb{\odot}$\fi}
\newcommand\rsun{\ifmmode{R\sb{\odot}}\else$R\sb{\odot}$\fi}
\newcommand\mearth{\ifmmode{M\sb{\oplus}}\else$M\sb{\oplus}$\fi}
\newcommand\rearth{\ifmmode{R\sb{\oplus}}\else$R\sb{\oplus}$\fi}

\renewcommand{\bm}[1]{{\mbox{{\boldmath$#1$}}}}	

\begin{document}

\title{Excitation of Inertial Modes in 3D Simulations of Rotating Convection in Planets and Stars}

\author{J. R. Fuentes}
\affiliation{\rm TAPIR, California Institute of Technology, Pasadena, CA 91125, USA}

\author{Ankit Barik}
\affiliation{\rm Department of Earth and Planetary Sciences,
The Johns Hopkins University, Baltimore, MD, USA}

\author{Jim Fuller}
\affiliation{\rm TAPIR, California Institute of Technology, Pasadena, CA 91125, USA}

\begin{abstract}
Thermal convection in rotating stars and planets drives anisotropic turbulence and  differential rotation, both capable of feeding energy into global oscillations. Using 3D simulations of rotating convection in spherical shells, we show that inertial modes--oscillations restored by the Coriolis force--emerge naturally in rotationally constrained turbulence,  without imposing any external forcing other than thermal/buoyancy driving. By varying the rotation rate at fixed Rayleigh number, we find that coherent modes appear only when the convective Rossby number, the ratio of the rotation period to the convective turnover time, falls below about one-half, where rotation dominates the dynamics. These modes are mostly retrograde in the rotating frame, equatorially symmetric, and confined to mid and high latitudes, with discrete frequencies well below twice the background rotation rate. At lower viscosities, or smaller Prandtl number, mode excitation becomes more efficient and a broader spectrum of inertial modes emerges. While the precise excitation mechanism remains uncertain, our results suggest that the modes are driven by instabilities due to differential rotation rather than stochastic forcing by convection. We conclude that similar inertial modes are likely to exist in the interiors of giant planets and stars, though their low frequencies will make them difficult to detect.
\end{abstract}


\section{Introduction}
\label{sec:Introduction}

Turbulence driven by thermal convection is a ubiquitous property of the atmospheres and interiors of planets and stars. The influence of the Coriolis force further organizes the flow and, through angular momentum transport, establishes differential rotation, which in turn feeds back on the convective motions.  This shear shapes the global circulation patterns, influencing magnetic field generation and the transport of heat and composition. Moreover, such large-scale shear flows could provide a source of free energy capable of exciting a variety of oscillation modes, either through direct instabilities or through nonlinear self-interactions of inertial waves, including triadic resonances \citep{Greenspan1968,Barik2018,Lin2021}. 

The most common oscillation modes in rotating flows are inertial modes--oscillations restored by the Coriolis force. They have been observed across a wide range of astrophysical and geophysical systems, from the solar convection zone \citep{Loptien2018,Gizon2021} and the convective cores of intermediate-mass stars \citep{Ouazzani2020,Saio2021}, to Earth's outer core \citep[e.g.,][]{Aldridge1987,Aldridge1988} and the oceans \citep[e.g.,][]{Fu1981,Niu2023,Khimchenko2024}. In these contexts, they have been shown to be valuable diagnostic tools for core rotation, differential rotation, as well as key contributors to ocean mixing, and atmosphere-ocean coupling. In addition, inertial modes play a central role in shaping the orbital and rotational evolution of stellar binaries, including star-planet and planet-moon systems \citep[e.g.,][]{Ogilvie2004,Wu2005,Barker2022}.

The properties of inertial modes are not fully-understood \citep[see, e.g.,][]{LeBars2015}. This is largely due to their strong sensitivity to the geometry and boundary conditions of the flow, and their coupling to nonlinear and dissipative processes, which are challenging to model and often unresolved in simulations. Yet, studies of rotating fluids in spherical geometry, ranging from early analyses of the linearized fluid equations \cite[e.g.,][]{Bryan1889,Rieutord1997,Zhang2001,Rieutord2001} to modern laboratory and numerical experiments of spherical Couette flow \citep[fluid confined in a spherical shell whose inner and outer boundaries rotate rigidly with different angular velocities, see, e.g.,][]{Triana2011,Matsui2011,Barik2018,Barik2024}, have revealed several robust features. To name a few, they are global waves with frequencies restricted to $|\omega| \leq 2\Omega$, where $\Omega$ is the angular frequency of the body or container, and non-axisymmetric modes drift azimuthally at a rate $\omega/m$, with $m$ the azimuthal wavenumber. In confined geometries, these waves focus along narrow shear layers or attractors, leading to enhanced shear and dissipation.

Regarding their excitation, it has been primarily attributed to wave–mean flow interactions. One proposed mechanism is over-reflection \citep{kelley2010}, in which incident inertial waves are amplified upon reflection from a shear layer, thereby extracting energy from the background differential rotation. In rotating bodies deformed by tidal interactions, inertial modes can be excited through the elliptical instability, a parametric instability in which pairs of inertial waves interact with the tidally induced strain field and extract energy from the tidal flow \citep[see, e.g.,][]{deVries2023}. Another is excitation at a "critical layer", often associated with the viscous shear layer that develops along the tangent cylinder in a rotating spherical shell \citep{Rieutord2012, Astoul2021}. In this case, if a region of the fluid co-rotates with the drift frequency of an inertial wave, resonant coupling can occur, leading to the amplification of the mode (i.e., critical layers correspond to regions of corotation resonance). Follow up studies of spherical Couette flow found no evidence for excitation at critical layers \citep{Hoff2016,Barik2018}, but did find excitation via shear instabilities of the axisymmetric background flow \citep{Barik2018}.

As described above, most studies of inertial modes have focused on spherical Couette setups, where the imposed differential rotation is simple and well controlled by the boundary conditions of the inner and outer shells. In more realistic astrophysical and geophysical systems, differential rotation emerges self-consistently from rotating convection. Inertial modes have been reported in only a handful of nonlinear simulations of rotating convection, notably in studies of Rossby waves in using spherical models of solar-like convection \citep{Bekki2022,Blume2024} and in simulations examining mode excitation in both full spheres and Cartesian geometries \citep{Lin2021,deVries2023}. These studies have shown that different families of inertial modes exist in both the radiative and convective zones of stars and planets, each with distinct spatial structures and properties that highlight the complex interplay between differential rotation and convection in exciting inertial modes.

In this work, we present a suite of 3D simulations of thermal convection in rotating spherical shells and analyze the resulting spatial and frequency spectra of the turbulence, along with the inertial modes excited in the system. We focus on how changes in the shell's rotation rate affect mode properties when the strength of convective thermal driving is kept constant. Section~\ref{sec:Numerical_Simulations} describes the model setup and numerical methods employed. Section~\ref{sec:results} presents our analysis and main findings. Finally, in Section~\ref{sec:discussion}, we summarize our results and discuss them in light of previous studies, as well as their implications for real astrophysical objects.

\section{Numerical Simulations}
\label{sec:Numerical_Simulations}

\subsection{Hydrodynamical Model}
\label{sec:nondimensionalization}

We simulate thermal convection in a 3D spherical shell of inner radius $r_i$ and outer radius $r_o$, initially rotating with constant angular frequency $\Omega_0\bm{\hat{z}}$. The gravity profile is $\bm{g} = g_0(r/r_o)$. The
spherical shell boundaries are assumed to be isothermal, impenetrable, and stress-free. The kinematic viscosity $\nu$ and the thermal diffusivity
$\kappa_T$ are assumed to be constant. For simplicity, we adopt the Boussinesq approximation \citep{Spiegel_Veronis_1960}, under which the flow is approximately incompressible, and density fluctuations are assumed to be small and linearly dependent on the temperature fluctuations,  $\rho/\rho_0 = - \alpha T$, where $\rho_0$ is the mean density of the fluid layer, $\alpha$ is the coefficient of thermal expansion, and $T$ is the the temperature perturbation relative to a constant reference temperature $T_0$.

We present the Boussinesq fluid equations in nondimensional form, using the shell depth $D = r_o - r_i$ as the unit of length and the viscous diffusion time $\tau_\nu = D^2/\nu$ as the unit of time. For the units of temperature, we use the temperature contrasts across the shell $\Delta T$. Finally, we adopt $\rho_0 D^2/\tau_\nu^2$ as the unit of pressure. After nondimensionalization, the radial domain goes from $r_i=3$ to $r_o = 4$, i.e., the fractional radius is $r_i/r_o = 0.75$, the shell depth is 1, and the resulting fluid equations become 

\begin{gather}
    \nabla \cdot \bm{u} = 0\, , \label{eq:div u}\\ 
 \dfrac{\partial\bm{u}}{\partial t} + \bm{u}\cdot \nabla \bm{u} - \nabla^2 \bm{u}   + \mathrm{Ek}^{-1}\bm{\hat{z}}\times \bm{u} = - \nabla P + \dfrac{\mathrm{Ra}}{\mathrm{Pr}}\dfrac{\bm{r}}{r_o}T~,
\\
    \dfrac{\partial T}{\partial t} + \bm{u}\cdot \nabla T  =  \dfrac{1}{\mathrm{Pr}} \nabla^2 T\, , \label{eq:T}
\end{gather}
where $\bm{r} = r\bm{\hat{r}}$, and $\bm{u}$ is the velocity field. Consistent with the Boussinesq approximation, we neglect the adiabatic temperature gradient in the thermal energy equation.

There are 3 dimensionless numbers that characterize the evolution of the flow. These are the thermal Rayleigh number, Prandtl number, and Ekman number, which can be expressed in terms of ratios of timescales
\begin{table*}
\centering
\small
\caption{Input and output parameters for each model in this study. With the exception of the last row, which has $\mathrm{Pr}=0.1$, all simulations use $\mathrm{Ra}=5\times10^6$ and $\mathrm{Pr}=1$. $\mathrm{Ra_c}$ denotes the critical Rayleigh number for the onset of convection at a given Ekman number. The first row corresponds to a non-rotating reference case. In our nondimensionalization, output Rossby numbers are defined as $\mathrm{Ro}_x = u_{x,\mathrm{rms}}\mathrm{Ek}$, where $x=r,\theta,\phi$, while the corresponding Reynolds numbers reduce to the dimensionless rms velocity, $\mathrm{Re}_x = u_{x,\mathrm{rms}}$.}
\label{table1}
\begin{tabular}{ccccccccc}
\hline
\multicolumn{3}{c}{Model inputs} & \multicolumn{6}{c}{Outputs} \\
\cline{1-3}\cline{4-9}
Ek & $\mathrm{Ro_c}$ & $\mathrm{Ra_c}$ &
$\mathrm{Ro}_r$ & $\mathrm{Ro}_\theta$ & $\mathrm{Ro}_\phi$ &
$\mathrm{Re}_r$ & $\mathrm{Re}_\theta$ & $\mathrm{Re}_\phi$ \\
\hline
$\infty$ & $\infty$ & $3.0\times10^3$ & $\infty$ & $\infty$ & $\infty$ & 324 & 378 & 392 \\
$6.33\times10^{-4}$ & 1.40 & $2.0\times10^{4}$ & 0.185 & 0.200 & 0.35 & 292 & 317 & 566 \\
$3.16\times10^{-4}$ & 0.70 & $4.7\times10^{4}$ & 0.077 & 0.081 & 0.48 & 244 & 257 & 1534 \\
$2.37\times10^{-4}$ & 0.53 & $6.7\times10^{4}$ & 0.048 & 0.053 & 0.40 & 203 & 223 & 1698 \\
$1.58\times10^{-4}$ & 0.35 & $1.1\times10^{5}$ & 0.021 & 0.034 & 0.11 & 134 & 214 & 720 \\
$1.18\times10^{-4}$ & 0.26 & $1.6\times10^{5}$ & 0.012 & 0.018 & 0.07 & 100 & 150 & 629 \\
$0.79\times10^{-4}$ & 0.17 & $2.6\times10^{5}$ & 0.005 & 0.007 & 0.03 & 61 & 91 & 440 \\
$2.50\times10^{-5}$ & 0.17 & $3.0\times10^{5}$ & 0.006 & 0.012 & 0.06 & 251 & 475 & 2292 \\
\hline
\end{tabular}
\end{table*}
\begin{equation}
\label{eq:dimensionless_numbers}
\mathrm{Ra} = \left(\frac{\tau_\nu}{\tau_{\rm ff}}\right)\left(\frac{\tau_\kappa}{\tau_{\rm ff}}\right)~, \quad \mathrm{Pr} = \frac{\tau_\kappa}{\tau_\nu}~, \quad \mathrm{Ek} = \left(\frac{\tau_{\Omega}}{\tau_\nu}\right)~,
\end{equation}
where $\tau_\kappa = D^2/\kappa_T$ is the thermal diffusion time across the shell, $\tau_\Omega = 1/2\Omega_0$ is the rotational timescale, and $\tau_{\rm ff} = \sqrt{D/\alpha g_o \Delta T }$ is the convective free-fall time across the shell, where $g_o$ is the acceleration of gravity at $r_o$. Another dimensionless number that characterize the flows is the Reynolds number, which in terms of the free fall velocity $u_{\rm ff} =\sqrt{\alpha g_o \Delta T D}$ can be written as a combination of the Rayleigh and Prandtl numbers, $\mathrm{Re} = u_{\rm ff}D/\nu = (\mathrm{Ra}/\mathrm{Pr})^{1/2}$.

Note that the ratio of the rotational timescale to the convective free-fall time defines the convective Rossby number $\mathrm{Ro_c}$

\begin{align}
\mathrm{Ro_c} = \dfrac{\tau_{\Omega}}{\tau_{\rm ff}} = \left(\dfrac{\mathrm{Ra}}{\mathrm{Pr}}\right)^{1/2} \mathrm{Ek}~.
\end{align}
Systems dominated by rotation have low values of $\mathrm{Ro_c}$, while flows that are relatively insensitive to rotation have high $\mathrm{Ro_c}$.  Our suite of simulations were conducted with the same Rayleigh and Prandtl numbers ($\mathrm{Ra} = 5\times 10^6$, $\mathrm{Pr} = 1$), but at different Ekman numbers ($\mathrm{Ek} \sim 7\times 10^{-5}$--$\infty$). In terms of the convective Rossby number, our simulations have  a broad dynamical range, with $\mathrm{Ro_c} \sim 0.17$--$\infty$, thereby allowing us to explore both rotationally constrained and rotationally unconstrained convective regimes. For comparison, we also conduct a single simulation using $\mathrm{Pr} = 0.1$, fixing $\mathrm{Ra}=5\times 10^6$ and $\mathrm{Ro_c}\approx 0.17$, so that $\mathrm{Ek}\sim 2.5\times 10^{-5}$. We present and analyze this simulation in detail in Section~\ref{sec:low_pr}.

All our rotating simulations are conducted at supercriticalities $\mathrm{Ra}/\mathrm{Ra_{c}(Ek)}$  in the rage 10--260, where $\mathrm{Ra_{c}(Ek)}$ is the critical Rayleigh number for the onset of convection at a given Ekman number. The values of the critical Rayleigh numbers $\mathrm{Ra_c}$ were obtained by interpolating values in the database of \cite{BarikEtAl2023} except for the $\mathrm{Pr} = 0,1$ case, which was computed using the linear code \texttt{Kore} \citep[https://github.com/repepo/kore,][]{BarikEtAl2023}. The non-rotating simulation has $\mathrm{Ra_{c}}\approx 3\times 10^3$, so that $\mathrm{Ra}/\mathrm{Ra_c}\approx 1660$. We also emphasize that realistic astrophysical parameters remain far beyond the reach of current computational capabilities. Nonetheless, by fixing 
$\mathrm{Ra}\gg 1,~ \mathrm{Ek}\ll 1$, and $\mathrm{Pr} \lesssim 1$, we ensure that our simulations remain qualitatively within the same dynamical regime as those relevant to gas and ice giant planets and stars. For more details on the input parameters and the actual flow parameters achieved in the simulations, see Table~\ref{table1}.


\subsection{Numerical Methods}

We time-evolve equations \eqref{eq:div u}--\eqref{eq:T} using the Dedalus pseudospectral solver \citep{Burns2020} version 3. The variables are represented in spherical harmonics for the angular directions and Chebyshev polynomials for the radial direction. The number of radial, latitudinal, and longitudinal coefficients in all the simulations are $(N_r,N_\theta,N_\phi) = (256,384,768)$, respectively. For time-stepping, we use a second order semi-implicit BDF scheme \citep[SBDF2,][]{wang_ruuth_2008}, where the linear and nonlinear terms are treated implicitly and explicitly, respectively. To ensure numerical stability, the size of the time steps is set by the Courant–Friedrichs–Lewy (CFL) condition, using a safety factor of 0.2 (based on trial and error). To prevent aliasing errors, we apply the ``3/2 rule'' in all directions when evaluating nonlinear terms. To start the simulations, we add small random-noise perturbations to the temperature field.

\section{Results} \label{sec:results}

\subsection{Morphology and Spatial Spectra of the Flow}
\begin{figure*}
    \centering
    \includegraphics[width=0.95\textwidth]{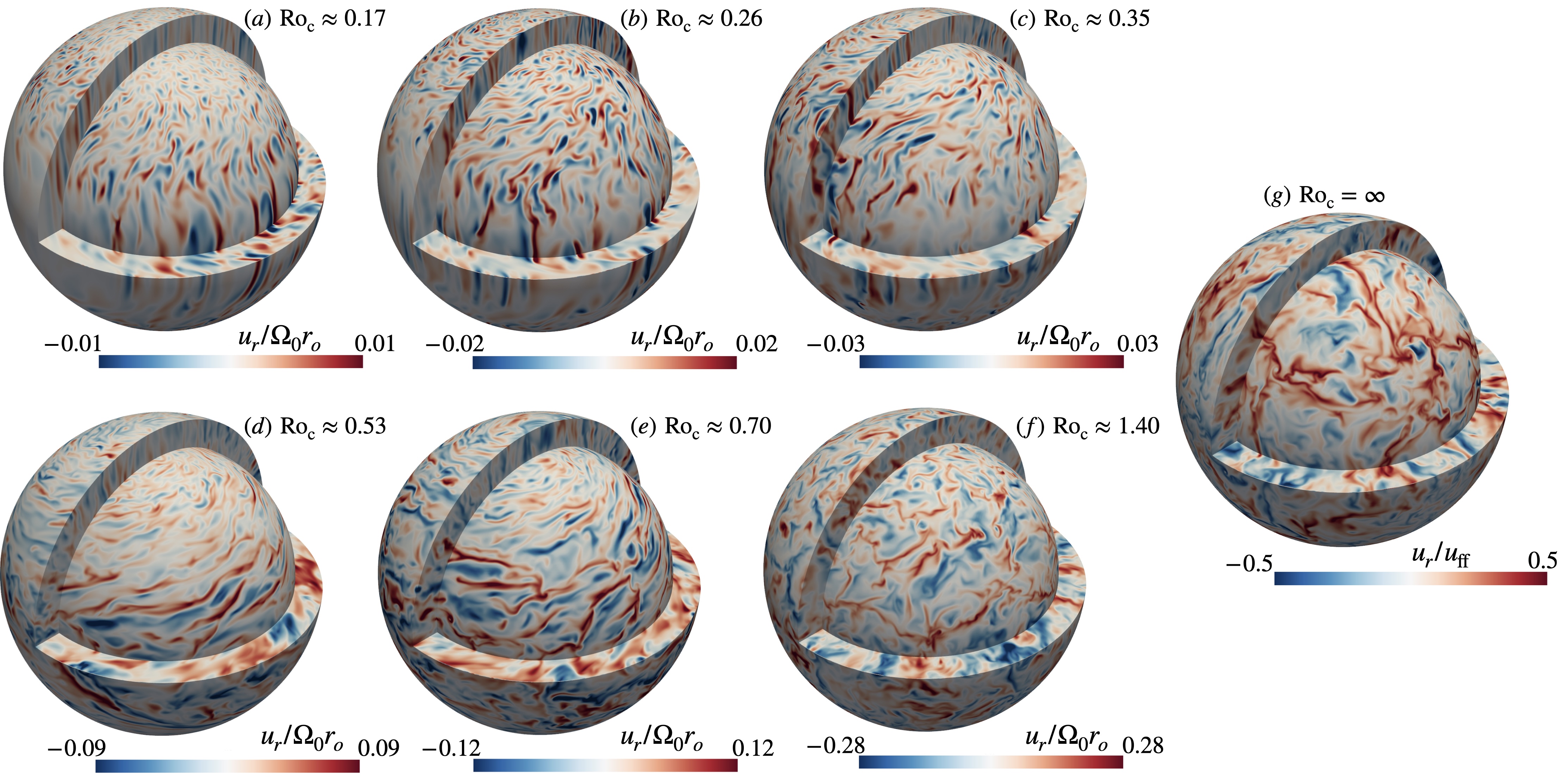}
    \caption{3D snapshots of the radial velocity $u_r$. The velocity in simulations of rotating flows is normalized by $\Omega_0 r_o$, while the non-rotating model is normalized by the free fall velocity $u_{\rm ff}$. Red and blue denote upflows and downflows, respectively. In the rotating cases, the flow exhibits markedly smaller, anisotropic spatial scales compared with the non-rotating case, where convection is more isotropic. All the simulations have the same Rayleigh number $\mathrm{Ra} = 5\times 10^6$ and Prandtl number $\mathrm{Pr} = 1$, while the Ekman number varies from $\mathrm{Ek}\sim 7\times 10^{-5}$ to $\mathrm{Ek} = \infty$ (non-rotating case). }
    \label{fig:u_r}
\end{figure*}

\begin{figure*}
    \centering
    \includegraphics[width=\textwidth]{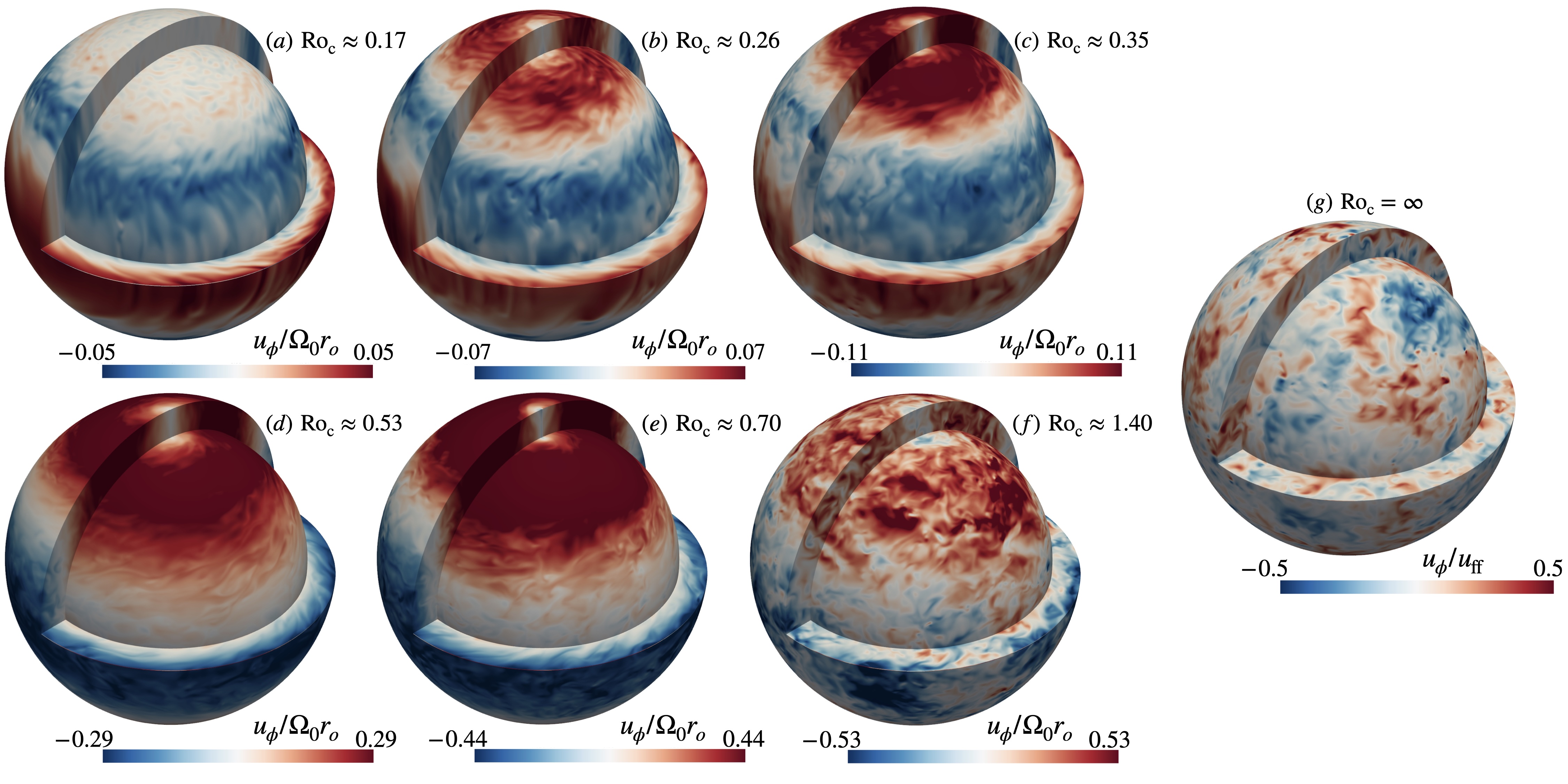}
\caption{3D snapshots of the azimuthal velocity $u_\phi$ for all simulations. The velocity in simulations of rotating flows is normalized by $\Omega_0 r_o$, while the non-rotating model is normalized by free fall velocity $u_{\rm ff}$. Red and blue denote prograde (eastward) and retrograde (wesward) direction, respectively. Simulations of small convective Rossby number $\mathrm{Ro_c}$ produce prograde equatorial jets, while simulations of large $\mathrm{Ro_c}$ yield  retrograde jets. No jets or differential rotation develop in the non-rotating case.}
    \label{fig:u_phi}
\end{figure*}

The radial velocity (Figure \ref{fig:u_r}) illustrates the morphological changes of the convective flow as rotation is varied. In the non-rotating case, the flow is more isotropic, with broad upwellings and downwellings and no preferred horizontal scale or alignment. As rotation increases (decreasing $\mathrm{Ro_c}$), the convective structures narrow and align with the rotation axis, forming the columnar patterns characteristic of rapidly rotating convection. At the smallest $\mathrm{Ro_c}$, the anisotropy between vertical and horizontal scales is most pronounced, and the flow is strongly constrained by the Coriolis force.

The corresponding azimuthal velocity (Figure~\ref{fig:u_phi}) reveal the large-scale zonal flows that develop in rotating convection.  These flows arise from Reynolds stresses generated in the convection zone \citep[see, e.g.,][]{Christensen2001, Busse2002,Aurnou2007}. For the most rapidly rotating cases ($\mathrm{Ro_c} \lesssim 0.5)$, a strong prograde (eastward) jet forms at the equator, flanked by retrograde (westward) flows at higher latitudes. As $\mathrm{Ro_c}$ increases, the amplitude and structure of these jets change: the equatorial prograde flow weakens, reverses sign, and gives way to a retrograde jet, with prograde flows at higher latitudes. This transition is consistent with previous studies linking the direction of equatorial jets to the convective Rossby number \citep[the well known solar to anti-solar differential rotation, see, e.g.,][]{Gastine2014, Camisassa2022}. In the absence of rotation, the zonal component lacks any coherent large-scale structure, and no mean (axisymmetric) differential rotation develops.

The influence of rotation on convection is further reflected in differences in the corresponding spatial power spectra. We compute these spectra using the SHTns package \citep{Schaeffer2013}, which employs a vector spherical harmonic decomposition of the velocity field at a given radius $r$ and time $t$
\begin{equation}
    \bm{u} = Q(\theta,\phi)\bm{\hat{r}} + r\nabla S(\theta,\phi) - \bm{r} \times \nabla T (\theta,\phi)~,
\end{equation}
where $Q, S$ and $T$ are the radial velocity component, and the spheroidal and toroidal scalar potentials, respectively. Expanding $Q$, $S$ and $T$ in spherical harmonics $Y_{\ell}^{m}(\theta,\phi)$ gives

\begin{align}
    \bm{u} &= \sum_{\ell = 0}^{\ell_{\rm max}}
    \sum_{m=-\ell}^{\ell} \Big[
        Q_{\ell}^{m} Y_{\ell}^{m}(\theta,\phi)\,\bm{\hat{r}} 
        + S_{\ell}^{m}\, r\nabla Y_{\ell}^{m}(\theta,\phi) \nonumber \\
    &\hspace{4.5em}
        - T_{\ell}^{m}\,\bm{r}\times \nabla Y_{\ell}^{m}(\theta,\phi)
    \Big]~,
\end{align}
where $Q_{\ell}^{m}$, $S_{\ell}^{m}$, and 
$T_{\ell}^{m}$ are the expansion coefficients of $Q$, $S$, and $T$, respectively.  

The total kinetic energy spectrum as a function of spherical harmonic degree $\ell$ is then computed as
\begin{equation}\label{eq:power}
\mathcal{P}(\ell,r,t) = \sum_{m\geq 0} C_m \left(|Q_{\ell}^{m}|^2 + \ell(\ell+1)(|S_{\ell}^m|^2 + |T_{\ell}^m|^2) \right)~,
\end{equation}
with $C_m = 1$ for $m=0$ and $C_m = 2$ for $m>0$, the latter accounting for the contributions from negative $m$.

\begin{figure*}
    \centering
    \includegraphics[width=\textwidth]{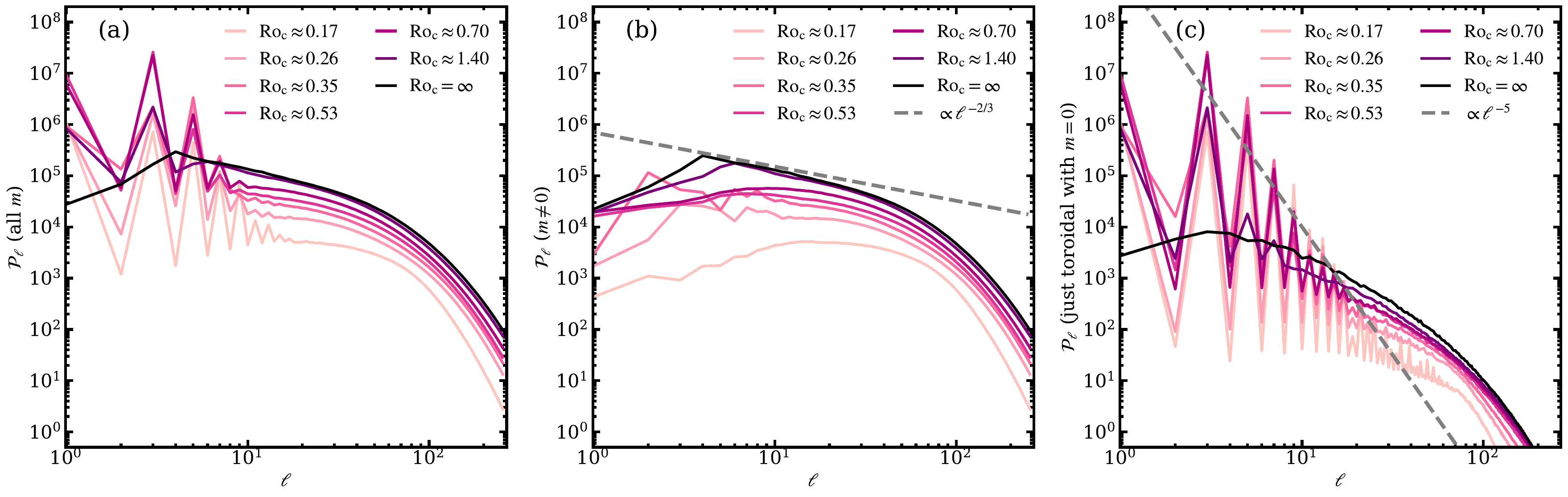}
    \caption{Velocity power spectra for all simulations. Flows were sampled at a radius $r=0.85r_o$ to construct the spectra, with other radii yielding essentially identical results. Panels (a) and (b) show the power associated with all azimuthal wavenumbers $m$ and with only the non-axisymmetric components ($m > 0$), respectively. Panel (c) shows the power of the zonal flow, i.e., the power considering only the toroidal component with the $m=0$ contribution in Equation~\eqref{eq:power}. 
    The dashed lines in panels (b) and (c) have a slope of $-2/3$ and $-5$ for comparison with Kolmogorov spectrum \citep{Kolmogorov1941} and with zonostropic turbulence \citep{Boning2023}, respectively. }
\label{fig:spatial_spectra}
\end{figure*}

Figure~\ref{fig:spatial_spectra} shows the power spectra for all our simulations, computed using the full set of azimuthal wavenumbers $m$ and, separately, using only the non-axisymmetric components ($m>0$), which represent the convective turbulence. All spectra were calculated at radius
$r = 0.85r_o$ and temporally averaged over a viscous diffusion time. We emphasize that the results are not sensitive to the choice of radius within the bulk of the shell, since the flow is nearly incompressible and density variations are small across the shell depth. In the absence of rotation, power simply decreases from large to small scales, as expected for non-rotating convection. With rotation, the distribution is more structured, axisymmetric motions ($m = 0$) dominate at low degrees and show peaks at odd $\ell$ values arising from toroidal equatorially symmetric zonal flows, while the convective part ($m>0$) grows with $\ell$ up to a critical value before declining in a way reminiscent of the non-rotating case.

When comparing the power distribution across different $\ell$ with theoretical expectations,  we find that for $\ell$ in the range 4--20 in our non-rotating model, the power exhibits a slope that is approximately consistent with the classic $-2/3$ scaling for homogeneous, isotropic turbulence \citep{Kolmogorov1941}\footnote{To avoid confusion, we emphasize that the Kolmogorov power spectrum is usually expressed in terms of the power per unit wavenumber $k$, i.e., $dE/dk \propto k^{-5/3}$. We can see that the ratio $P_\ell/\ell \propto \ell^{-5/3}$ as expected for Kolmogorov's scaling.}
, and it steepens significantly for $\ell \gtrsim 50$, where the dissipation range begins. The slope for the rotating models is much shallower. This is not surprising and has been noted by previous work \citep[see, e.g.,][]{Featherstone2016}. The main reason is that turbulent flows in rapidly rotating convection are highly anisotropic. 

It is worth noting that the power associated with the zonal differential rotation (i.e., the toroidal, $m=0$ contribution in Equation~\ref{eq:power}), which dominates the spectral peaks, behaves quite differently from that of convective turbulence. Our numerical results indicate that the zonal power follows an $\ell^{-5}$ scaling (Figure~\ref{fig:spatial_spectra}d), a scaling that also appears in solar differential rotation from SDO/HMI observations, in MHD simulations of jet formation in Saturn and hot Jupiters, and even in laboratory experiments \citep{Yadav2020,Lemasquerier2023, Boning2023,Boning2024}.

As argued by \citet{Rhines1975} in the context of $\beta$-plane turbulence, the characteristic jet width $d$ and speed $U$ in planetary atmospheres satisfy $d\sim \sqrt{U R/2\Omega\sin\theta}$, where $\Omega$ is the rotation rate and $\theta$ the colatitude. Adopting this scaling and extrapolating it to spherical geometry, we identify the jet width with spherical harmonic degree, $d\sim \pi R/\ell$, and estimate the jet velocity from the definition of the zonal energy spectrum, $U \sim \sqrt{2\ell P_{\ell,\rm zonal}}$. Since the zonal spectrum is dominated by low-$\ell$ modes, one obtains the scaling $P_{\ell, \rm zonal} \propto \ell^{-5}$.

\begin{figure*}
    \centering
    \includegraphics[width=\textwidth]{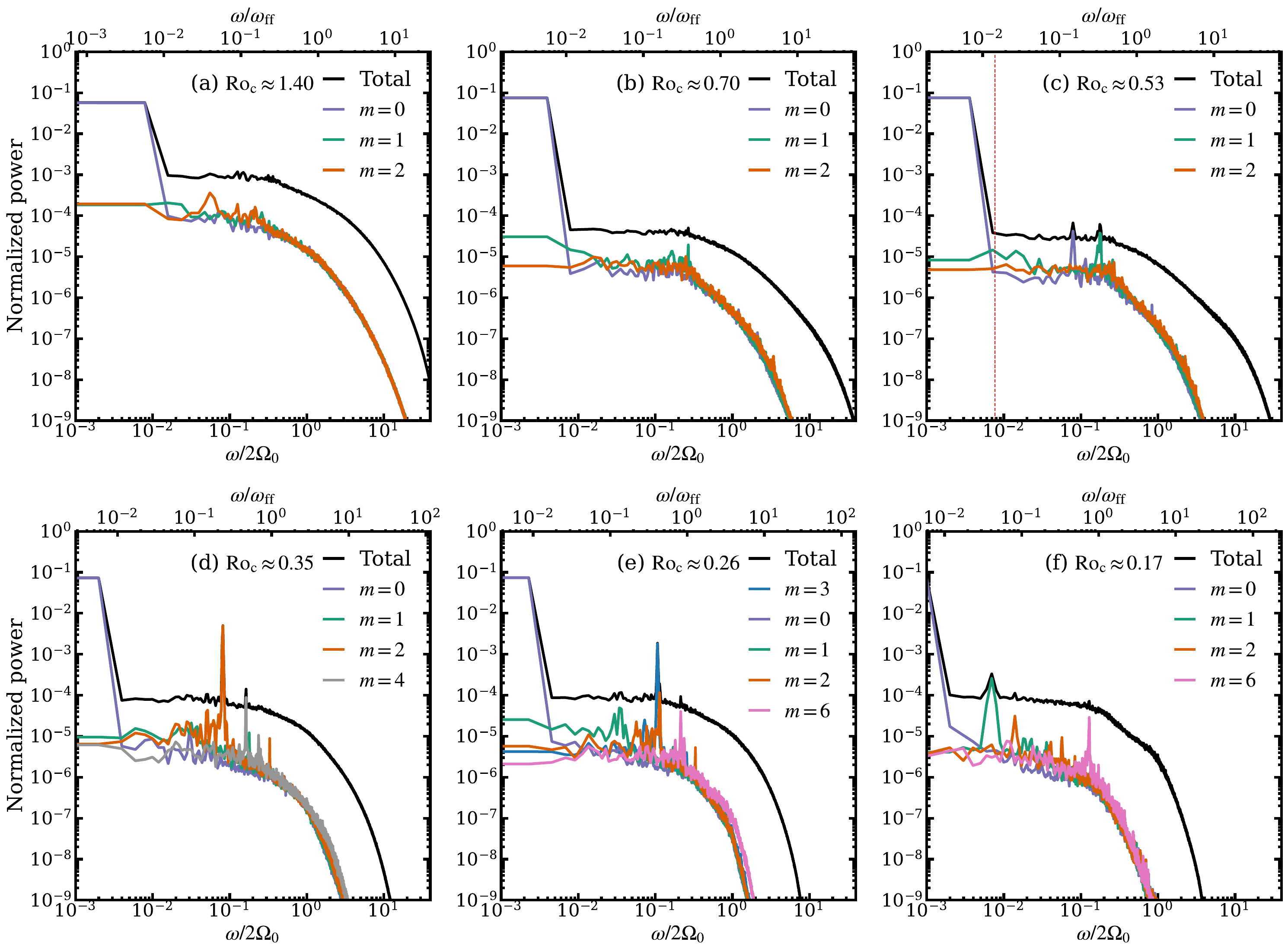}
    \caption{Power spectra of the kinetic energy at $r=0.85r_o$, as a function of frequency for all rotating models. We show the frequency normalized to $2\Omega_0$, and to the convective frequency $\omega_{\rm ff}$. The power is normalized by the total over all frequencies, $\ell$, and $m$. Black curves show the sum over all $m$ and $\ell$, while colored curves indicate the contributions from all $\ell$ but individual $m$, allowing identification of the modes responsible for the observed peaks.}
    \label{fig:frequency_spectra}
\end{figure*}

\begin{figure*}
    \centering
    \includegraphics[width=0.935\textwidth]{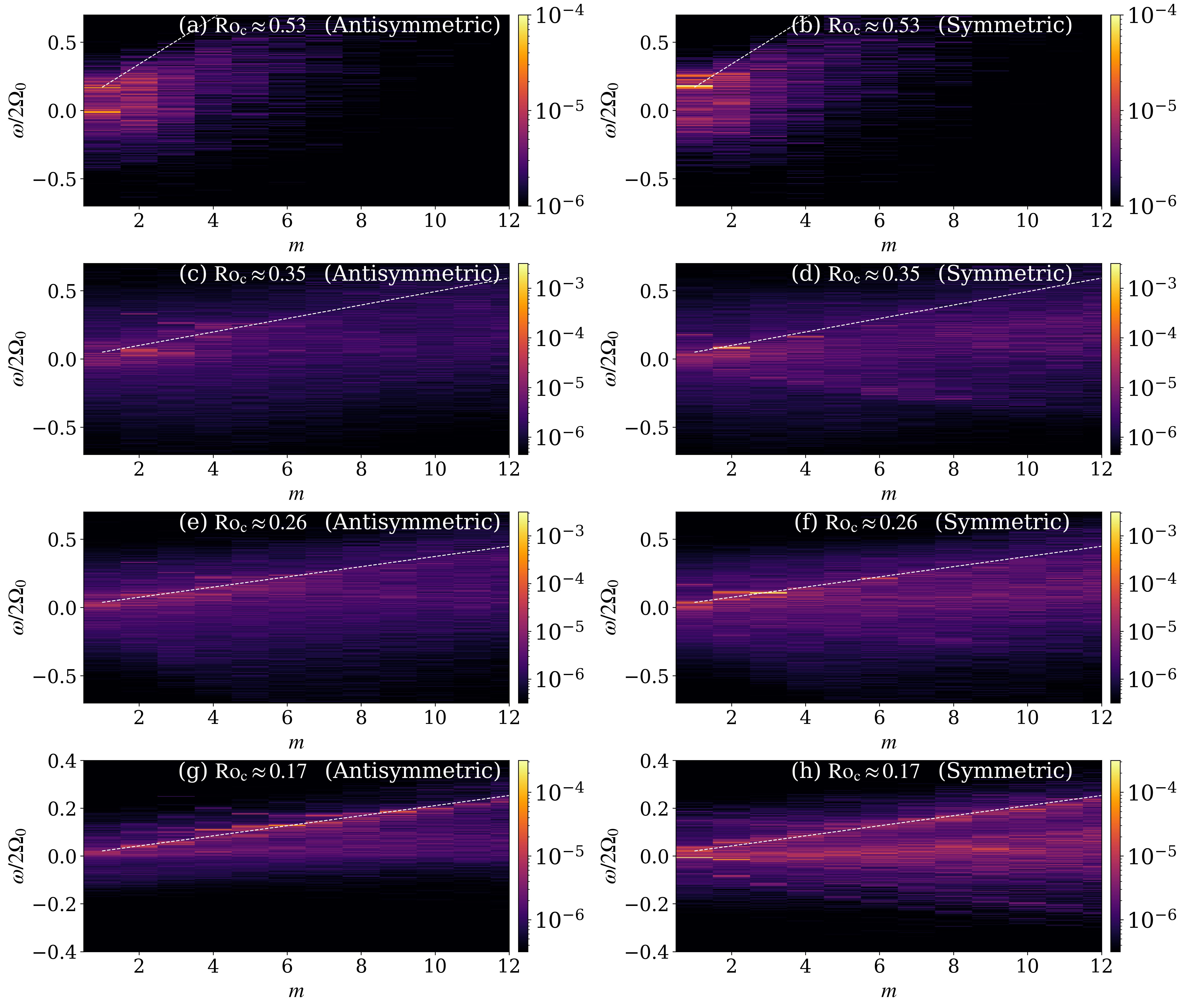}
    \caption{Power spectra of the kinetic energy at $r=0.85r_o$, as a function of azimuthal order $m$ and dimensionless temporal frequency $\omega/2\Omega_0$ in the rotating frame.  The left panels shows the antisymmetric contribution to the power (summing only over the signals with $\ell - m$ odd), while the right panels shows the symmetric contribution to the power (summing only the signals with $\ell - m$ even). The dashed lines indicate $\omega / 2\Omega_0 \simeq m \Delta\Omega / 2\Omega_0$, where $\Delta\Omega = |\min(\Omega)|$ represents the maximum retrograde (negative) shear of the mean flow in the rotating frame. Results are presented for simulations with $\mathrm{Ro_c} \leq 0.53$.}
    \label{fig:2d_spec}
\end{figure*}

\subsection{Frequency Spectra}

A direct way to identify the excitation of oscillation modes in a simulation is to analyze the flow in frequency space.  We proceed in much the same way as when examining the flow across spatial scales, by constructing the power spectrum as a function of spherical harmonic degree $\ell$, order $m$, and frequency $\omega$. The starting point is recording of the velocity field at a fixed radius over a sequence of equally spaced time intervals. At each snapshot, every component of the velocity is expanded in spherical harmonics, yielding a time series of complex spectral coefficients $u^{r,\theta,\phi}_{\ell m}(t)$ for each ($\ell,m$) mode. Applying a temporal Fourier transform to each coefficient reveals its oscillatory content, and the squared amplitudes $|\hat{u}^{r,\theta,\phi}_{\ell m}(\omega)|^2$ measure the kinetic energy associated with each mode at a given frequency. Summing over all the spherical harmonic degree $\ell$, order $m$, and all velocity components, produces the total kinetic energy spectrum per frequency bin,  $\mathcal{P}(\omega) \approx (dE/d\omega) \Delta \omega$,

\begin{equation}
\mathcal{P}(\omega) = \sum^{\ell_{\mathrm{max}}}_{\ell=0}\sum^{\ell}_{m=0}\left(|\hat{u}^{r}_{\ell m}(\omega)|^2 + |\hat{u}^{\theta}_{\ell m}(\omega)|^2 + |\hat{u}^{\phi}_{\ell m}(\omega)|^2\right) C_{\ell m}~,
\end{equation}
where $C_{\ell m} = 1$ for $m=0$, and  $C_{\ell m} = 2$ for $m > 0$ (where the factor of 2 accounts for contributions from negative $m$). 

Figure~\ref{fig:frequency_spectra} shows the power in the frequency spectrum of the flow at $r=0.85 r_o$, for all the simulations with the exception of the non-rotating case, which is omitted as it shows no notable spectral features. Distinct coherent peaks in the spectrum reveal the excitation of modes, and the broader structure reflects the background of convective turbulence. The total frequency spectra refer to the ones summed over all $\ell$ and $m$, while ones for specific $m$ are summed over all $\ell$ for that specific order. Note that each spectrum contains an $m=0$ contribution that is much larger at very low frequencies. This corresponds to the ``zero-frequency'' mean flows associated with differential rotation.

Similarly to the non-rotating case, at relatively high convective Rossby numbers ($\mathrm{Ro_c}\approx 1.4$ and 0.7 in panels a and b, respectively), the spectra remain largely featureless. Clear peaks emerge only for $\mathrm{Ro_c} \lesssim 0.53$, all at frequencies below twice the rotation rate ($\omega/2\Omega_0 < 1$), consistent with inertial modes. At $\mathrm{Ro_c} \approx 0.53$ (panel c), an axisymmetric ($m=0$) mode is clearly present alongside an $m=1$ mode. 

As the convective Rossby number decreases to 0.35 (panel d), a prominent $m=2$ mode appears, accompanied by a weaker $m=4$ mode whose frequency is twice that of $m=2$, indicating a non-linear self-interaction that resemble the ones described in \cite{Barik2018} for the spherical Couette system.
At $\mathrm{Ro_c} \approx 0.26$ (panel e), the spectrum is dominated by an $m = 3$ mode, followed by an $m=2$ signal at a nearby frequency, and low amplitude $m = 1$ and $m=6$ modes. The frequency of the $m=6$ mode is exactly twice that of the $m=3$ mode, indicating that it arises from a self-interaction of the lower mode. At the lowest Rossby number in the suite, $\mathrm{Ro_c} \approx 0.17$ (panel f), we find three small-amplitude modes corresponding to $m=1$, $m=2$, and $m=6$, with the $m=2$ mode having a frequency roughly twice that of the $m=1$ mode. We emphasize that the amplitude of the excited modes depends on both the damping and the excitation mechanisms, which in our simulations are likely dominated by viscous dissipation and nonlinear interactions with convective motions (i.e. an effective turbulent viscosity), as well as by the degree of differential rotation. Although rotation weakens convection, particularly at low Rossby numbers, all of our rotating simulations exhibit strong differential rotation. However, we do not find any clear correlation between the mode amplitude and the strength of the differential rotation.

For comparison, each panel also shows the frequency of the spectrum normalized to the free-fall frequency $\omega_{\rm ff} \sim D/u_{\rm ff}$. We can see that the power spectrum is fairly flat for $\omega/\omega_{\rm ff} \lesssim 1$, and it generally falls steeply for $\omega/\omega_{\rm ff} \gtrsim 1$, as expected. However, it does not clearly exhibit the $\mathcal{P}(\omega) \propto \omega^{-2}$ spectrum expected for Kolmogorov turbulence with $\omega > \omega_{\rm ff}$ \citep{goldreich:90,goldreich:94}. We might expect different slopes in three regimes, with the lowest frequencies being $\omega < \omega_{\rm ff}$, the intermediate frequency regime in $\omega_{\rm ff} < \omega < 2 \Omega_0$, and the high-frequency regime being $\omega > 2 \Omega_0$. Our $\mathrm{Ro_c}\approx 0.17$ simulation appears to have a shallower slope to the power spectrum in the intermediate frequency regime than the high-frequency regime. For the other simulations, the separation in scales between $\omega_{\rm ff}$ and $2 \Omega$ is too small to see any clear differences.

To provide a broader perspective, Figure~\ref{fig:2d_spec} shows the frequency spectra as a function of azimuthal wavenumber $m$, including both positive and negative frequencies, and separated into equatorially symmetric and antisymmetric components. A symmetric component is characterized by $u_r(r,\theta,\phi) = u_r(r,\pi-\theta,\phi),~u_\theta(r,\theta,\phi)= -u_\theta(r,\pi-\theta,\phi),~u_\phi(r,\theta,\phi)= u_\phi(r,\pi-\theta,\phi)$ and thus, requires summing over degrees such that $\ell - m$ is even for the $r$ and $\phi$ velocity components and odd for the $\theta$-component for each $m$. The exact opposite holds for antisymmetric components. In addition to confirming the coherent peaks identified in Figure~\ref{fig:frequency_spectra}, the 2D spectra also reveal that most of the power of the strongest modes resides in the symmetric component. With the exception of the $m=1$ mode at $\mathrm{Ro_c} \approx 0.17$, the strongest modes have positive frequencies. In the sign convention used here, each mode varies as $e^{i(m\phi + \omega t)}$, so that a constant phase condition gives $d\phi/dt = -\omega/m$. Because the calculations are performed in the rotating frame and we take $m>0$, modes with positive (negative) frequencies correspond to retrograde (prograde) propagation relative to the rotating frame. Thus, only the $m=1$ mode at $\mathrm{Ro_c} \approx 0.17$ drifts prograde, while all others drift retrograde.

Another interesting feature of the 2D spectra is the presence of ``ridges'' where the convective power is concentrated along lines of $\omega/2\Omega_0 \propto m$, most clearly at positive frequencies. These ridges have approximately constant pattern speed, $\omega/m$, suggesting they arise from low-frequency waves advected by differential rotation. The dashed lines correspond to $\omega/2\Omega_0 \sim m \Delta \Omega/2\Omega_0$, with $\Delta \Omega$ = $|\min(\Omega)|$, i.e., the maximum negative (retrograde) shear of the flow in the rotating frame. This alignment indicates that the waves along the positive-frequency ridges are advected by the retrograde zonal jets present in the simulations.

\subsection{Observed Wave Structures}
\begin{figure}
    \centering
\includegraphics[width=\columnwidth]{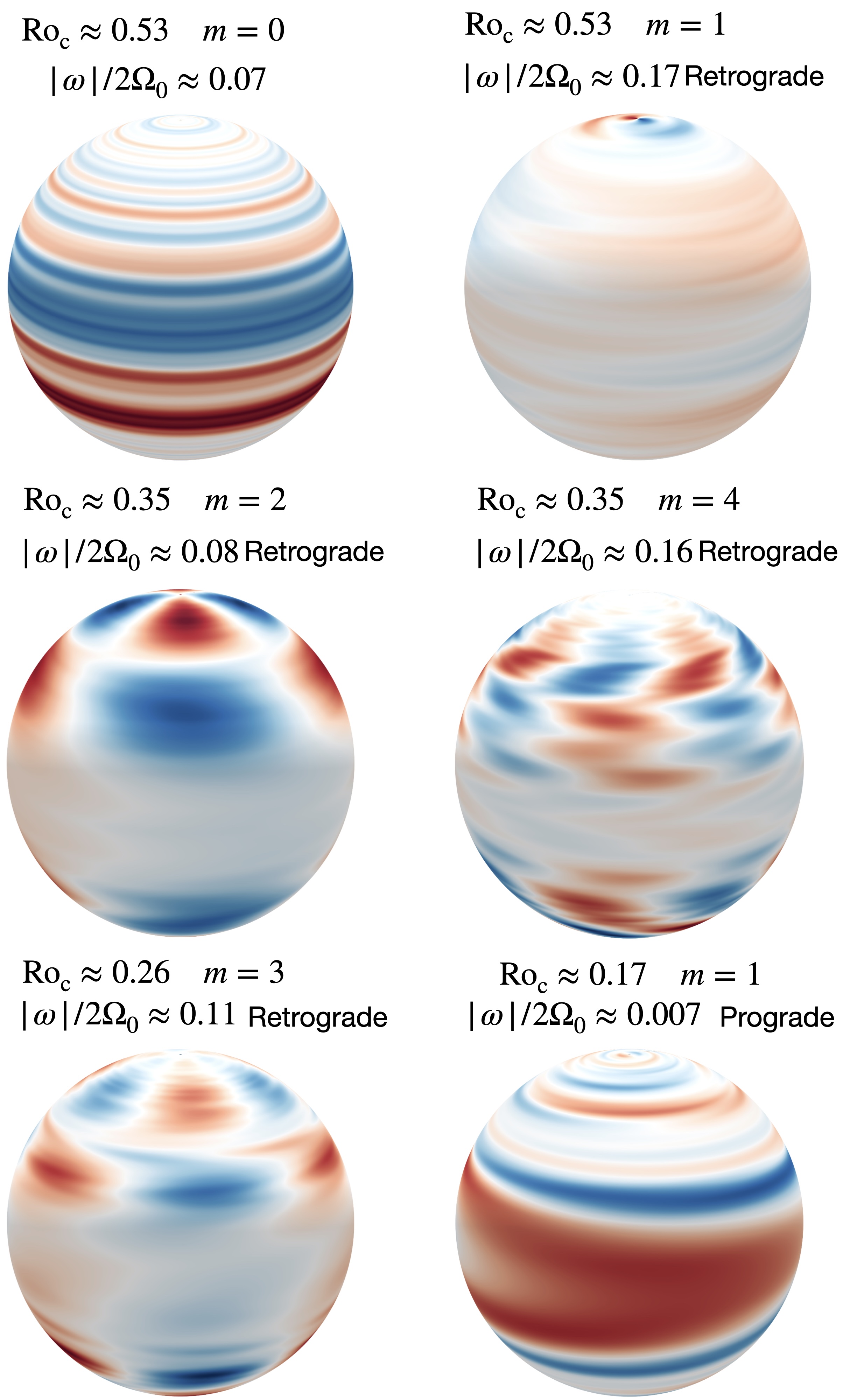}
    \caption{Azimuthal velocity component ($u_\phi$) projected on the spherical surface, filtered by the azimuthal wavenumber $m$ and frequency $\omega$ corresponding to the excited mode. Similarly to Figure~\ref{fig:u_phi}, red and blue means positive and negative, respectively. The color scale is normalized to the maximum value. The $m=1$ mode for $\mathrm{Ro_c}\approx 0.53$ is confined to the polar regions.}
    \label{fig:modes}
\end{figure}
Figure~\ref{fig:modes} shows the morphology of several modes excited in the simulations. We find that $u_\phi$ provides the strongest contribution to the power spectrum in these modes. Thus, we filter the zonal component of the velocity in space and time, isolating the azimuthal wavenumber $m$ and the mode frequency. The presence of differential rotation and contamination from convective motions makes it difficult to identify them in terms of pure eigenmodes. 
Nevertheless, these projections confirm that all the inertial modes in our simulations, except for the $m=0$ mode at $\mathrm{Ro_c}\approx 0.53$, exhibit symmetry about the equator, as also suggested by the dominance of the symmetric component in Figure~\ref{fig:2d_spec}.

\subsection{Internal Shear Layers and Mode Attractors}

Another useful measure is the meridional distribution of the zonal kinetic energy, which reveals the presence of internal shear layers  \citep{StewartsonRickard1969,Kerswell1995,Rieutord2001}. These shear layers are aligned with the characteristic rays of inertial waves, which propagate at critical co-latitudes defined by  $\cos\theta_c = \omega/2\Omega_0$, forming  rays tilted by an angle  $\pi/2 - \theta_c$ with respect to the rotation axis. In a spherical shell, the corresponding boundary-value problem becomes ill-posed, producing singularities along these paths \citep{StewartsonRickard1969,Kerswell1995}.
Viscosity removes these singularities and produces narrow regions of strong shear along the attractor paths. 

Because nearly all the inertial modes excited in our simulations are non-axisymmetric, any azimuthal averaging would erase their signal from the energy profiles. Therefore, we follow the same approach as in Figure~\ref{fig:modes}, i.e., we filter the azimuthal velocity component by the wavenumber $m$ of the dominant mode, and then compute $u^2_\phi(r,\theta,\phi_0)$ at fixed azimuthal angle $\phi_0$. We note that frequency filtering is not feasible here, as these profiles require full 3D spatial information, while only a limited number of time snapshots were stored for computational reasons (frequency filtering would require storing the 3D data at high temporal cadence).

\begin{figure}
    \centering
    \includegraphics[width=\columnwidth]{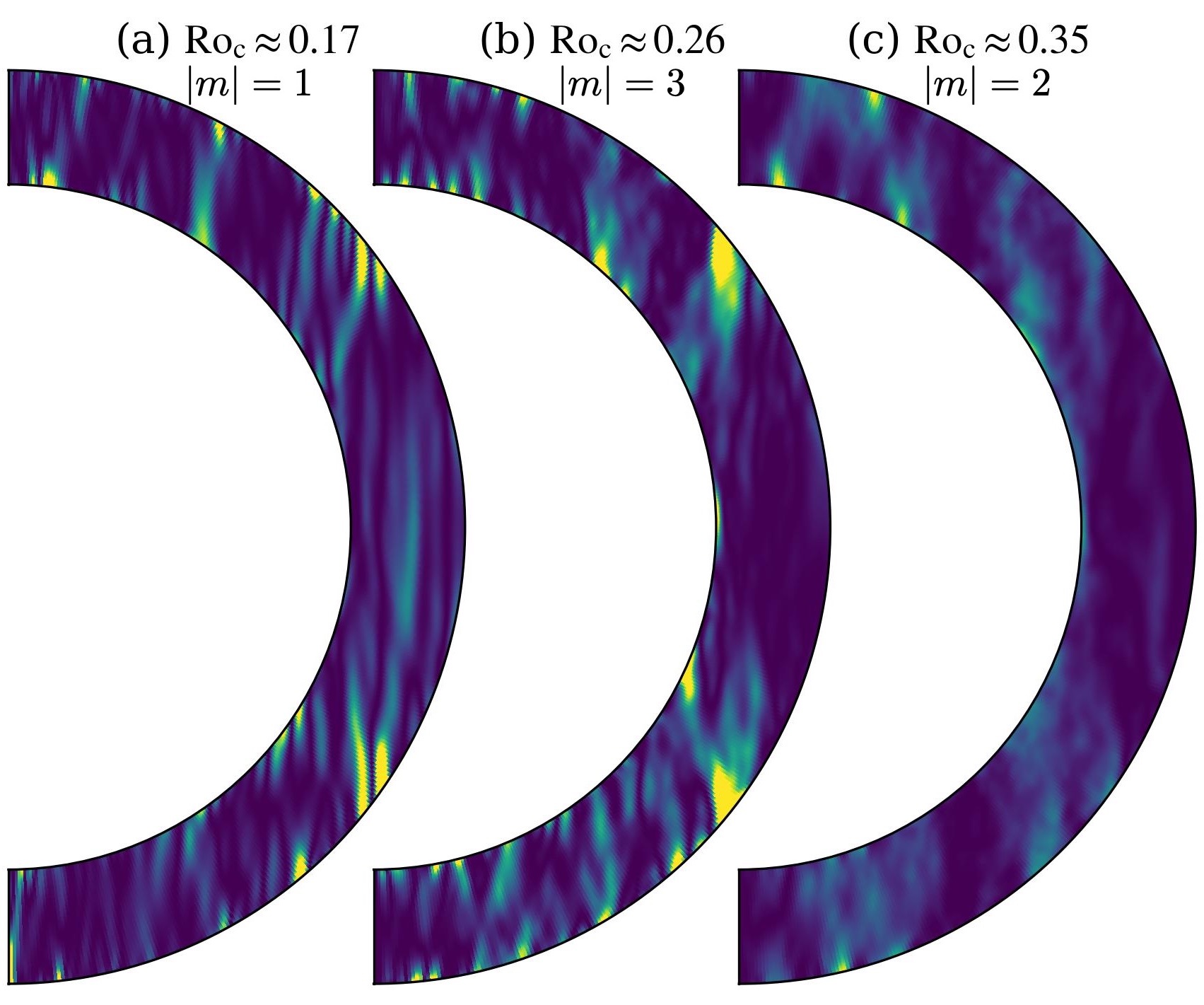}
    \caption{Meridional profiles of the zonal kinetic energy, filtered by the azimuthal wavenumber $m$ of the dominant modes in each simulation. Dark and light colors indicate regions of low and high energy, respectively. The profiles in panels (a)–(c) are shown at azimuthal angles $\phi_0 = \pi/8,~\pi/24,~\pi/16$, chosen to highlight the clearest mode signatures. 
Internal shear layers resembling wave attractors are clearly seen as reflecting rays at high latitudes.}
    \label{fig:shear}
\end{figure}
Figure~\ref{fig:shear} shows the meridional distribution of zonal kinetic energy for the cases at $\mathrm{Ro_c}\approx 0.17$, 0.26, and 0.35, which are the most rapidly rotating cases  and whose modes have the largest peaks in the frequency spectra. Internal shear layers are visible as ``ray reflections'', especially at low $\mathrm{Ro_c}$. For $\mathrm{Ro_c} \approx 0.17$ (panel a), the flow is dominated by an $m=1$ mode with a low frequency, $|\omega|/2\Omega_0\approx 0.007$, producing nearly vertical shear layers aligned with the rotation axis. At higher $\mathrm{Ro_c}$ (panels b and c), the dominant modes exhibit larger frequencies, $|\omega|/2\Omega_0\sim 0.1$, resulting in shear layers inclined at greater angles to the axis. Note that the shear layers appear more diffuse for the case with $\mathrm{Ro_c}\approx 0.35$. This can be attributed to a larger viscous spreading, as the Ekman number in that model is twice as large as in the case with $\mathrm{Ro_c}\approx 0.17$. 

\subsection{Low Prandtl number}\label{sec:low_pr}

Given the large thermal and radiative diffusivities of astrophysical fluids, the Prandtl number is typically much less than unity ($\mathrm{Pr} \ll 1$). It is therefore of interest to compare our results with cases at lower $\mathrm{Pr}$. Owing to computational limitations, we perform a single additional simulation with $\mathrm{Pr} = 0.1$, fixing $\mathrm{Ra} = 5\times10^6$ and $\mathrm{Ro_c} \approx 0.17$, which corresponds to an Ekman number $\mathrm{Ek} \approx 2.5\times10^{-5}$.

\begin{figure}
    \centering
    \includegraphics[width=\columnwidth]{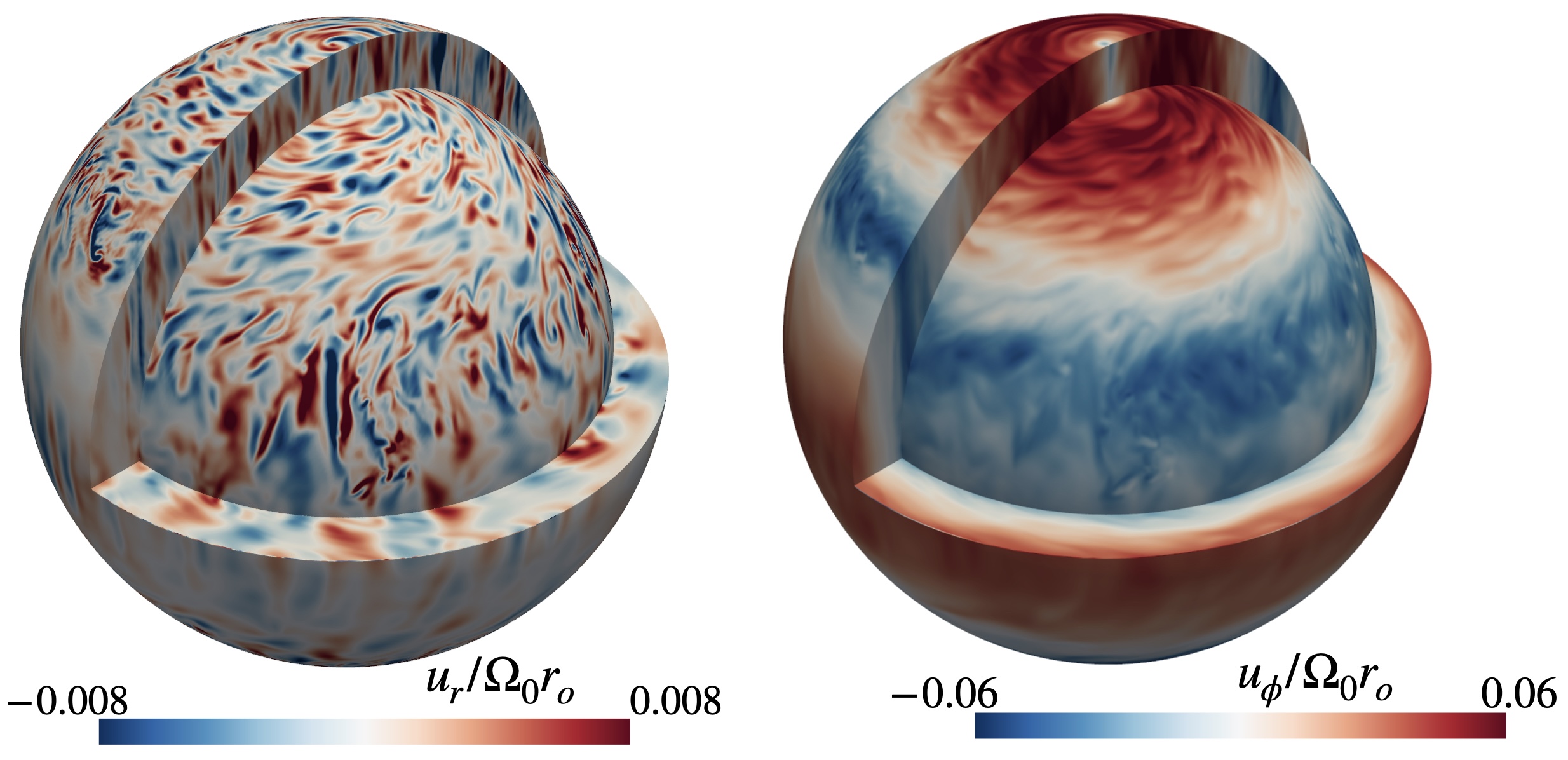}
    \caption{3D snapshots of the radial $u_r$ and azimuthal velocity $u_\phi$ (left and right panel, respectively) for the simulation using $\mathrm{Pr} = 0.1$ and $\mathrm{Ro_c}\approx 0.17$. The velocities are normalized by $\Omega_0 r_o$. Unlike the case using $\mathrm{Pr} = 1$ and $\mathrm{Ro_c}\approx 0.17$, where the flow velocities are weak at high latitude, the convective flow and the differential rotation for the  $\mathrm{Pr} = 0.1$ are vigorous at at latitudes. }
    \label{fig:low_pr}
\end{figure}

\begin{figure*}
    \centering
    \includegraphics[width=0.3\textwidth]{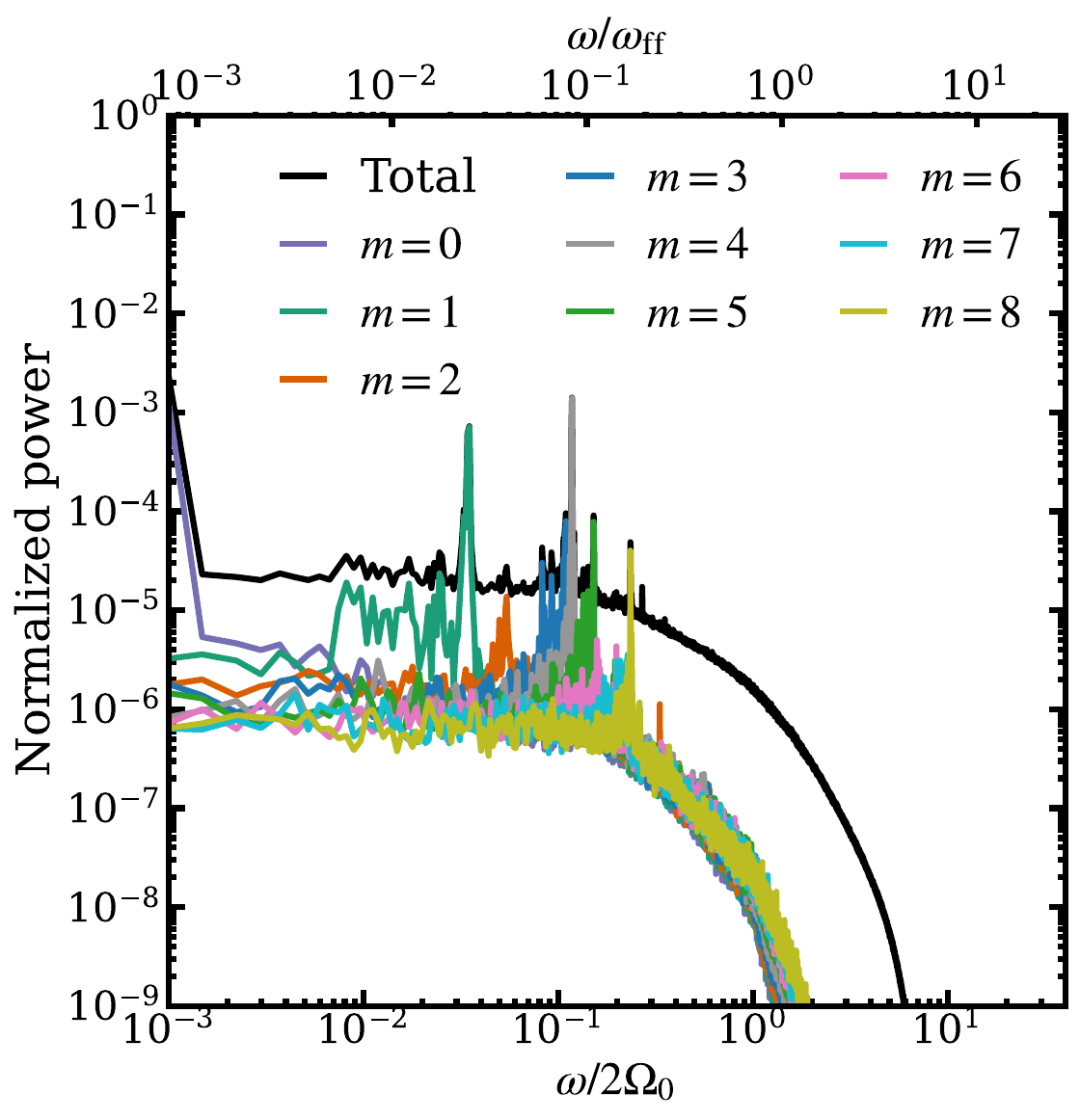}\includegraphics[width=0.7\textwidth]{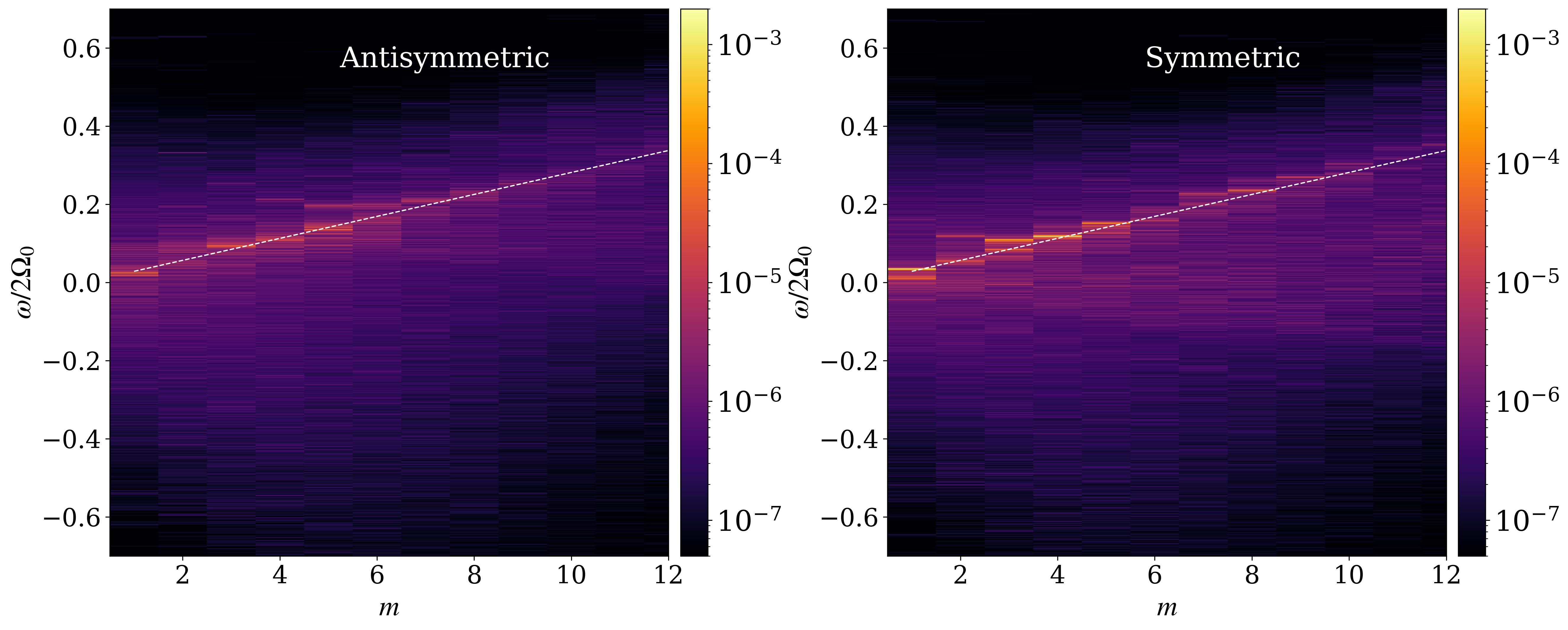}
    \caption{Power spectrum of the kinetic energy at $r=0.85r_o$, for the simulation using $\mathrm{Pr} = 0.1$ and $\mathrm{Ro_c}\approx 0.17$. Results are shown as a function of only frequency (left panel), and as a function of frequency and $m$ order, separating contributions from equatorially antisymmetric and symmetric motions (middle and right panels). The normalization and the meaning of the several lines in these  plots are the same as in Figures~\ref{fig:frequency_spectra} and $\ref{fig:2d_spec}$.}
    \label{fig:low_pr_spectrum}
\end{figure*}

\begin{figure}
    \centering
\includegraphics[width=\columnwidth]{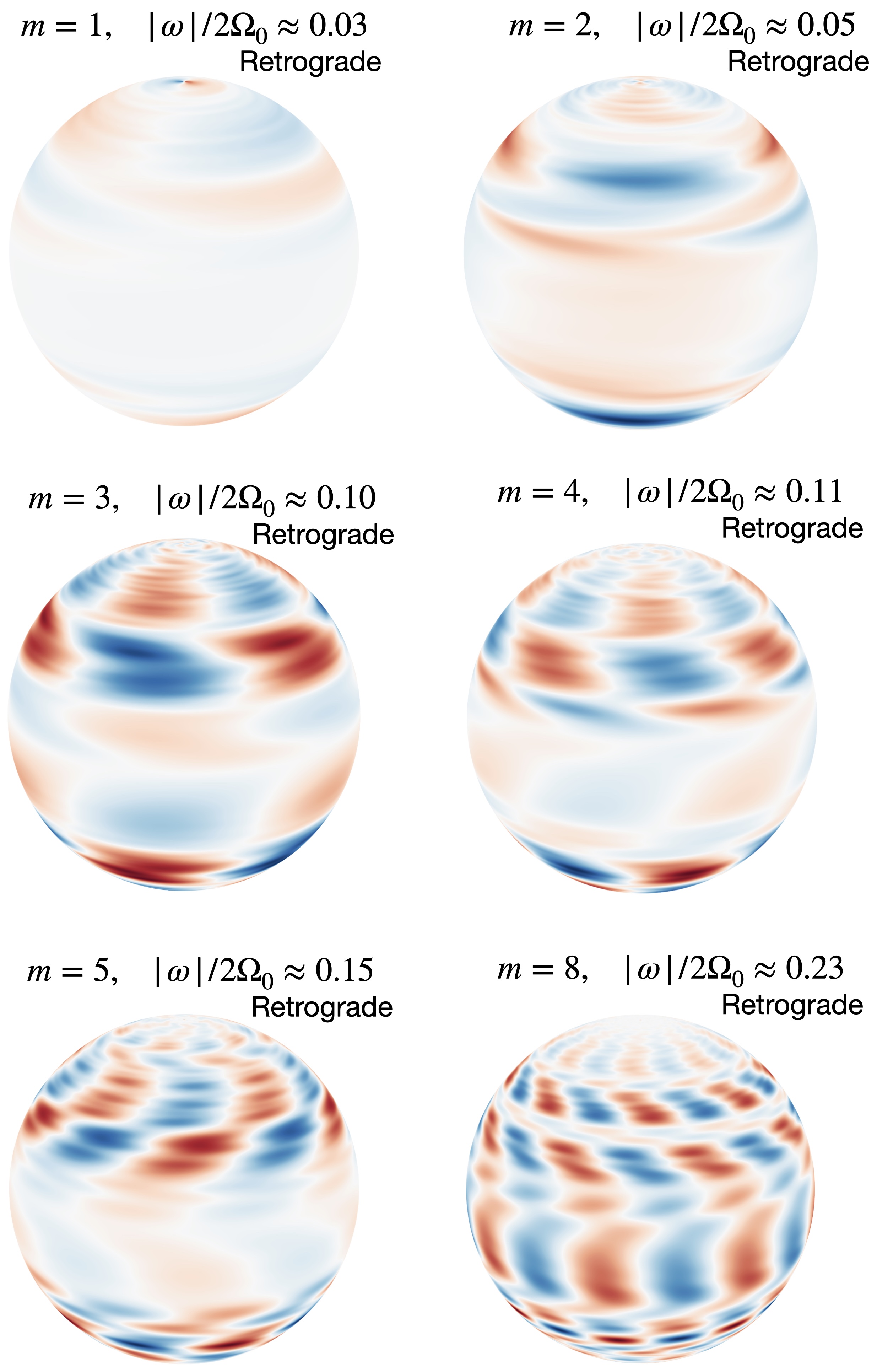}
    \caption{Same as Figure~\ref{fig:modes} but the inertial modes excited in the simulation using $\mathrm{Pr} = 0.1$, $\mathrm{Ro_c}\approx 0.17$.}
    \label{fig:modes_low_pr}
\end{figure}

We find several differences relative to the model using $\mathrm{Pr} = 1$. For example, because the effective buoyancy scales as $\mathrm{Ra}/\mathrm{Pr}$, the flow velocities are much larger in the $\mathrm{Pr} = 0.1$ simulation (compare the last two rows of Table~\ref{table1}). Another difference is the strength of both the convective flow and the differential rotation. While the $\mathrm{Pr}= 1$ case shows weak flows at high latitudes (see panels a of Figures~\ref{fig:u_r} and \ref{fig:u_phi}), the $\mathrm{Pr} = 0.1$ model exhibits more vigorous flows at all latitudes (see the radial and azimuthal components of the velocity in Figure~\ref{fig:low_pr}).

The frequency spectrum of the modes also shows remarkable differences with respect to the $\mathrm{Pr} = 1$ case. As shown by Figure~\ref{fig:low_pr_spectrum}, mode excitation is more efficient at low Pr, with several large peaks appearing over the range $m=1$--8. These inertial modes are mostly equatorially symmetric and drift retrograde in the rotating frame. In contrast, the $\mathrm{Pr} = 1$ case exhibits much weaker excitation, with only three modes of very small amplitude (see panel f of Figure~\ref{fig:frequency_spectra}), even though the convective Rossby number $\mathrm{Ro_c} \approx 0.17$ remains the same in both simulations. As discussed previously, the dominant $m=1$ mode in this case is equatorially symmetric but prograde in the rotating frame. Finally, another difference is the spatial location of the modes. While the low amplitude $m=1$ mode observed for $\mathrm{Pr} = 1$ is confined to the equatorial region (see Figure~\ref{fig:modes}), the modes for $\mathrm{Pr} = 0.1$ (including the $m=1$ mode) have most of their power from mid to high latitudes (see Figure~\ref{fig:modes_low_pr}).
Though the onset of convection at low $\mathrm{Pr}$ occurs in the form of oscillatory quasi-geostrophic inertial-waves \citep[QGIWs,][]{Zhang2004}, our simulation at $\mathrm{Pr}=0.1$ is sufficiently supercritical that we do not expect these QGIWs to play a significant role in the nonlinear dynamics. Nevertheless, a more systematic investigation of their influence will be addressed in future work.

\section{Summary and Discussion}\label{sec:discussion}

In this work, we have studied inertial modes using 3D simulations of thermal convection in rotating spherical shells. A key distinction from previous studies is that the differential rotation in our simulations develops naturally from the interaction between convection and rotation, rather than being imposed through boundary forcing as in spherical Couette setups. We also varied the convective Rossby number $\mathrm{Ro_c} = (\mathrm{Ra}/\mathrm{Pr})^{1/2}\mathrm{Ek}$ between 0.17 and 1.4, keeping the Rayleigh number fixed at $\mathrm{Ra} = 5\times 10^6$ and varying the Ekman number $\mathrm{Ek}$ between $7\times 10^{-5}$ and $6\times 10^{-4}$ in simulations of $\mathrm{Pr} = 1$. This choice produced flows with distinct differential rotation profiles that lead to different properties of the excited oscillation modes. We also conducted a single simulation at a lower Prandtl number, using $\mathrm{Ra} = 5\times 10^6$, $\mathrm{Ek}\sim 2.5\times 10^{-5}$, and $\mathrm{Pr} = 0.1$, i.e., the convective Rossby number was kept fixed to  $\mathrm{Ro_c}\approx 0.17$, for comparison with the $\mathrm{Pr} = 1$ model.

In the spatial domain, the kinetic energy power spectra of the flow resemble those reported in previous numerical simulations of rotating convection \citep{Featherstone2016,Hindman2020}, with energy dominated by axisymmetric motions that produce strong peaks at odd spherical harmonic degrees $\ell$ (see Figure~\ref{fig:spatial_spectra}). Interestingly, the zonal component of the kinetic energy power spectra exhibits an $\ell^{-5}$ scaling, as observed in the spectra of the zonal flows of the gas and ice giants \citep[e.g.,][]{Sanchez2000,Sukoriansky2002}, argued for zonostrophic turbulence \citep[see the recent preprint by][]{Boning2024,Cabanes2024}.

In the frequency domain, the low frequency range of the power spectra is dominated by the $m=0$ contribution of the differential rotation. At higher frequencies, the spectra show an overall flat distribution with superimposed peaks, followed by a rapid decay for $\omega/\omega_{\rm ff} > 1$, i.e., for frequencies larger than the convective frequency (see Figures~\ref{fig:frequency_spectra} and \ref{fig:low_pr_spectrum}). These coherent peaks, which correspond to inertial modes excited within the flow, are only observed in cases with $\mathrm{Ro_c} \lesssim 0.53$, while they are absent at larger Rossby numbers. This is expected since the amount of differential rotation increases for more rapidly rotating flows. However, the amplitude of the peaks does not appear to correlate with the Rossby number and  depends on the growth rate and saturation of each mode, influenced by both the damping and excitation mechanisms. We emphasize that the absence of inertial modes at larger Rossby numbers is specific to modes excited self-consistently by rotating convection in our setup, and may not hold for other excitation mechanisms or different choice of parameters.
The Prandtl number nevertheless appears to play a more important role in mode excitation, since the $\mathrm{Pr} = 0.1$ simulation exhibits a richer and more vigorous spectrum of inertial modes than any of the simulations with $\mathrm{Pr} = 1$.

For the most part, the inertial modes in our simulations are nonaxisymmetric, retrograde in the rotating frame, symmetric about the equator, and span azimuthal orders $1 \leq m \leq 8$, with frequencies $\omega / 2\Omega_0 \lesssim 0.2$ (see Figure~\ref{fig:2d_spec} and \ref{fig:low_pr_spectrum}). Their structures are complex, likely due to the effects of turbulent convection and differential rotation, and show that nearly all modes are confined to mid and high latitudes. This suggests that their excitation may be linked to shear instabilities within the differentially rotating convection zone. Further, their zonal energy distributions reveal internal shear layers that resemble mode attractors, particularly at high latitudes near the locations where the zonal flow changes sign (see Figure~\ref{fig:shear}).

An exception is the small amplitude $m=1$ at $\mathrm{Ro_c}\approx 0.17$ and $\mathrm{Pr} = 1$, which is localized near the equator and drifts in the prograde direction with respect to the rotating frame (see Figures~\ref{fig:modes} and \ref{fig:modes_low_pr}). Other exceptions are the $m=8$ mode in the $\mathrm{Pr} = 0.1$ simulation (which has power at low latitudes) and the $m=3$ mode in the same simulation (which is equatorially anti-symmetric).

We also found a single axisymmetric mode ($m=0$) of frequency $\omega/2\Omega_0 \approx 0.1$ in the simulation with $\mathrm{Ro_c}\approx 0.53$. The origin of this mode is uncertain, since shear instabilities are not expected to excite axisymmetric modes, and the differential rotation in this simulation is not unstable to the centrifugal instability \citep{Rayleigh1917,Bayly1988}, which is the usual mechanism that gives rise to axisymmetric modes. Nevertheless, we believe that this mode is real since its signal corresponds to a distinct and sharp peak in the frequency spectrum, that is clearly separated from the ``zero frequency'' power associated with the differential rotation (see Figure~\ref{fig:frequency_spectra}c).

\cite{Zhang2001} found a special class of equatorially symmetric inertial modes which strongly satisfy the Taylor-Proudman constraint \citep[i.e., if the pressure gradient and the Coriolis force balance each other, see][]{Proudman1916,Taylor1917}. In such scenario, the modes are symmetric with respect to the equator, their frequencies satisfy $-i\omega~\bm{u} = -\nabla P/\rho - 2\Omega_0 \bm{\hat{z}}\times \bm{u} \approx 0$, and in turn, they are expected to be nearly geostrophic ($|\omega|/\Omega_0\ll 1$). Those modes are referred to as ``slow'' inertial modes \citep{Barik2018}. 
In contrast, other equatorially symmetric as well as antisymmetric modes that do not satisfy the Taylor–Proudman constraint can have much higher frequencies \citep[see also][]{Zhang2001,Wicht2014}. The modes observed in our simulations are predominantly equatorially symmetric but have frequencies $\omega/2\Omega_0 \sim 0.1$, suggesting that they are not close to the geostrophic limit.

Another difference is about the excitation mechanism. In general, the inertial modes observed in simulations and laboratory experiments of spherical Couette flows have been attributed to shear instabilities. It is likely that many of the modes in our simulations are excited simultaneously by both convection and differential rotation, with only a few reaching large enough amplitudes to produce a coherent and prominent peak in the frequency spectra. In all our simulations we found the presence of ``critical layers'', regions where the drift frequency of the mode co-rotates with the local angular velocity of the flow. These regions are typically where non-axisymmetric inertial modes are expected to extract energy from the background shear flow and grow in amplitude \citep{Rieutord2012, Astoul2021}. However, \citet{Baruteau2013} analyzed the interaction between inertial modes and differential rotation in a spherical shell, finding that for a cylindrical rotation profile of the form $\Omega(s) \propto 1 + Cs^2$, modes are absorbed and damped at the co-rotation layer. Our simulations also produce cylindrical rotation, but with profiles far more complex than a simple quadratic in $s$, and the modes persist. Their persistence implies that either the simple profile result does not generalize, or another mechanism offsets corotation damping. 

Our simulations lack an obvious oscillatory mechanism--such as libration, precession, or tides--that would typically generate and sustain inertial modes \citep{LeBars2015}. The only external energy input to the system is thermal, which drives convection and establishes the differential rotation. 
How this energy is transferred into the inertial modes, and sustains them once excited, remains an open question. As noted earlier, the excitation and selection of inertial modes are long-standing problems, even in systems such as spherical Couette flow, where such modes have been observed for decades \citep{kelley2010}. Shear instabilities of the mean zonal flow have been proposed as a possible mechanism in several contexts \citep{Garaud2001,Barik2018}, and this may also apply to our simulations, which maintain a persistent differential rotation. Nonetheless, we emphasize that the detailed pathway of excitation and the process by which certain modes are preferentially amplified remain to be determined. 

Many of the same dynamics that occur in our simulations may occur in planets or stars. The differential rotation that develops due to convection is an obvious example that has been studied quite extensively. Our simulations do not appear to exhibit Rossby modes and other classes of inertial modes that are observed in simulations of the Sun (e.g., \citealt{Blume2024}), but the reason is not clear. One possibility is the lack of magnetic fields and density stratification in our models, both of which were included in  \citealt{Blume2024}.

Although the amplitudes of the modes excited stochastically by convection depend on the frequency spectrum of the turbulent flow \citep{goldreich:90}, it typically leads to the excitation of a broad set of modes whose amplitudes reflect the underlying convective power spectrum. In contrast, in our simulations we observe only a small number of discrete inertial modes reaching large, long-lived, and phase-coherent amplitudes. This behavior is more naturally explained by linear excitation through a shear instability as seen in previous studies \citep{Barik2018,Souza-Gomes2025}.
In the Sun, inertial modes have recently been detected via helioseismology (see \citealt{Gizon2024} and references therein), and some are excited stochastically \citep{Philidet2023} while others are excited by a linear instability \citep{Bekki2022b}. Those authors attribute the excitation mechanism to baroclinic instability, but we believe this is unlikely in our case because our boundary conditions enforce nearly barotropic flow, as evidenced by the cylindrical flow patterns that develop. In any case, there is no obvious reason that similar mode excitation would not occur in real planets and stars.

If inertial modes are excited by differential rotation in gaseous planets, they may be detectable. The high-latitude unstable $m=1$ modes in the Sun have velocity amplitudes of $\sim 5$ m/s \citep{Gizon2024}, roughly 10 times larger than the largest amplitude p-modes. Based on our simulation results, the unstable inertial modes have angular frequencies $\omega \lesssim 0.1 \Omega_0$, corresponding to oscillation periods $\gtrsim$ 4 days (measured in a co-rotating frame) in Jupiter and Saturn. However, the frequency in an inertial frame would be $\sigma \approx m \Omega_0$, corresponding to periods $P \approx 10/m \, {\rm hours}$ for Jupiter and Saturn. These low frequencies may be difficult to detect with ground-based methods, e.g., radial velocity monitoring \citep{Gaulme2011}. Doppler tracking of an orbiting satellite (Parisi et al. in preparation) may offer a possibility, but the small gravity perturbations produced by inertial modes make this seem unlikely.

In fact, unstable inertial modes and/or r-modes may have already been observed in Saturn via ring seismology. \cite{Hedman2014} have detected several $m=3$ waves in Saturn's rings whose intertial-frame pattern frequencies are very close to Saturn's rotation rate, i.e., they have very low frequencies in Saturn's co-rotating frame. \cite{Friedson2023} argued that these could be caused by r-modes in Saturn's stably stratified interior, and showed that such modes could be unstable due to Saturn's differential rotation. The inertial modes in our simulation are similar, except that they are confined to the convective layers rather than stably stratified layers. In this scenario, inertial and/or r-modes of other $m$ would likely also be excited within Saturn, but only the $m=3$ modes have the right inertial-frame frequencies to launch waves at Lindblad resonances in Saturn's C-ring.

To our knowledge, these types of inertial modes have not been observed in stars, even though they may be present. Inertial modes in convective cores of $\gamma$-Doradus stars have been detected \citep{Ouazzani2020,Saio2021}, but these are really gravito-inertial modes excited by near-surface effects \citep{Neiner2012,Guzik2000}. Thousands of solar-type stars and red giants exhibit p~modes stochastically excited by convection, but low-frequency inertial modes have not been detected (as far as we are aware), perhaps because they produce very small photometric and radial velocity modulations. There are many ``hump-and-spike" stars that appear to exhibit a dense spectrum of r-modes at pattern frequencies $\omega/m$ slightly less than the rotation rate \citep{Saio2018}. It may possible that some of the observed modes in these stars are caused by low-frequency inertial modes excited in the convective core, which would be Doppler shifted by rotation to pattern frequencies near the stellar rotation rate.

We have made several assumptions and approximations that should be relaxed in future work. For example, we adopted the  Boussinesq approximation \citep{Spiegel_Veronis_1960}, which restricts the flow to be roughly incompressible, i.e., the density is approximately constant. Planets and stars have large density variations across their interiors and are highly compressible. As has been shown by several studies \citep{LockitchFriedman1999,Wu2005,Mukhopadhyay2025}, density stratification can change the power spectra at different radii, and affect the eigenfunctions of the modes, particularly those of non-toroidal modes with significant radial motions. Density variations also change the morphology of the convective flow, creating a strong asymmetry with fast, narrow downflows and slow, wider upflows.

We have also excluded stably-stratified regions in the fluid. Such regions are predicted to exist in the cores of giant planets \citep[e.g.,][]{Fuller2014,Mankovich2021}, and could harbor a much richer wave spectrum composed of gravito-inertial modes, thereby influencing the orbital dynamics of planet-moon systems through tidal interactions \citep[e.g.,][]{Pontin2024,Dhouib2024}. We have also excluded magnetic fields in our simulations whose presence adds the Lorentz force as an additional restoring force giving rise to magneto-Coriolis modes \citep{Finlay2008}.

In order to investigate inertial modes across different differential rotation profiles, we have focused our study on varying the convective Rossby number, $\mathrm{Ro_c}$, while keeping the Prandtl number fixed at unity. Yet, as shown in Section~\ref{sec:low_pr}, lowering the Prandtl number to $\mathrm{Pr} = 0.1$ leads to a far richer and more vigorous spectrum of inertial modes, with amplitudes exceeding those seen in the $\mathrm{Pr} = 1$ cases. This suggests that the Prandtl number may play a more fundamental role in mode excitation than previously appreciated. Future simulations that push toward the astrophysically relevant regime $\mathrm{Pr} \ll 1$ would therefore be of great value for understanding inertial modes in planets and stars.

\begin{acknowledgements}
We thank Adrian Barker for his careful review that helped us improve our manuscript. J.R.F. is supported by the Sherman Fairchild Postdoctoral (Burke) Fellowship and the Presidential Fellowship at Caltech, as well as NASA Solar System Workings grant 80NSSC24K0927. JF is grateful for support through the Caltech-JPL  President's and Director's Research \& Development Fund Program.
\end{acknowledgements}






\bibliography{references}{}
\bibliographystyle{aasjournal}

\end{document}